\documentclass[10pt]{article}

\usepackage[utf8]{inputenc}
\usepackage[T1]{fontenc}
\usepackage{mathptmx}
\usepackage[scaled=0.92]{helvet}
\usepackage{amsmath,amssymb}
\usepackage{graphicx}
\usepackage[table]{xcolor}
\usepackage{booktabs}
\usepackage{array}
\usepackage{multirow}
\usepackage{caption}
\usepackage{enumitem}
\usepackage{titlesec}
\usepackage[numbers,sort]{natbib}
\usepackage[a4paper,top=2.45cm,bottom=2.45cm,left=2.3cm,right=2.3cm]{geometry}
\usepackage[hidelinks]{hyperref}
\usepackage{xurl}
\usepackage{placeins}

\definecolor{ncehead}{HTML}{27374D}
\definecolor{nceblue}{HTML}{3F6FA3}
\definecolor{ncepink}{HTML}{C75B70}
\definecolor{ncegreen}{HTML}{3E8E5E}
\definecolor{ncepurple}{HTML}{7E5AA2}
\definecolor{rowpastel}{HTML}{EAF2FB}
\definecolor{rowpastelb}{HTML}{F3EEF8}
\definecolor{headpastel}{HTML}{D7E6F4}

\titleformat{\section}{\normalfont\large\bfseries}{\thesection.}{0.5em}{}
\titleformat{\subsection}{\normalfont\normalsize\bfseries}{\thesubsection.}{0.5em}{}
\titlespacing*{\section}{0pt}{1.4ex plus 0.6ex}{0.7ex}
\titlespacing*{\subsection}{0pt}{1.1ex plus 0.5ex}{0.5ex}

\hypersetup{colorlinks=true,linkcolor=nceblue,citecolor=ncegreen,urlcolor=nceblue}

\newcommand{\code}[1]{\texttt{\small #1}}

\begin{document}
\pagestyle{plain}

\begin{center}
{\LARGE\bfseries An Open-Source LFSR-Based Stochastic Leaky Integrate-and-Fire Neuron in SkyWater 130\,nm: Design, Stochastic Characterisation, and Rate Coding}\\[10pt]
{\large Poornima Kumaresan and Santhosh Sivasubramani\textsuperscript{*}}\\[4pt]
{\itshape Intrinsic Lab, Centre for Sensors, Instrumentation and Cyber-Physical System Engineering (SeNSE), Indian Institute of Technology Delhi, New Delhi 110016, India}\\[3pt]
\textsuperscript{*}Corresponding author. E-mail: \href{mailto:ssivasub@iitd.ac.in}{ssivasub@iitd.ac.in}, \href{mailto:ragansanthosh@ieee.org}{ragansanthosh@ieee.org}
\end{center}

\vspace{0.4em}
\noindent\textbf{Abstract.}
Stochastic spiking neurons trade exact arithmetic for controlled randomness, which lowers area and tolerates input noise, and they are a useful primitive for event-driven edge hardware and for interfacing with emerging analogue synapses. This paper presents the design and characterisation of a compact, configurable stochastic leaky integrate-and-fire neuron implemented as standard-cell CMOS on the SkyWater 130\,nm process and released openly. A 16-bit configurable-polynomial linear-feedback shift register drives an eight-entry programmable activation table whose output sets a Bernoulli firing probability, and a saturating 16-bit leaky integrator with a programmable threshold and a refractory period of zero to seven cycles produces the spike train. All parameters are written through a sixteen-register serial interface, and the neuron runs either from parallel inputs or entirely from the register file. We characterise the block from a model checked bit-exact against the register-transfer code: the shift-register period is maximal at 65535 states for a maximal-length polynomial and short at 63 states for the shipped default, the eight-bit comparison value is uniform over the full period, and the per-entry firing probability equals the table value divided by 256. We also characterise a property that a system-level model of the source would not expose: although the underlying maximal-length sequence is ideal, the comparator output is serially correlated at short lags, with a negative lobe near lag eight, because the compared byte shifts by one bit each cycle, and subsampling every sixteen cycles restores whiteness. Rate-coding sweeps show monotonic control of the output rate by the input weight and by the threshold, and the refractory period caps the rate at one spike per refractory-plus-one cycles. The neuron occupies about 10,600 square micrometres of standard cells, places at 70 per cent utilisation on a single Tiny Tapeout tile, meets the 50\,MHz timing constraint with positive setup and hold margin, and is verified by eighteen directed cocotb tests at register-transfer and gate level. All results are from simulation and the open implementation flow; no fabricated silicon is reported. The neuron is an openly released companion to a four-block neuromorphic suite reported separately \cite{suite2026}.

\vspace{0.5em}

\vspace{0.2em}

\section{Introduction}\label{sec:intro}
Edge devices that classify and detect close to their sensors operate under power and area budgets that conventional clocked arithmetic meets poorly. Neuromorphic hardware addresses this regime by encoding information as sparse spikes and by computing where the data reside, which lowers the arithmetic and the data movement that dominate energy use \cite{davies2018loihi,merolla2014}. Within this style of computation, a neuron that fires probabilistically as a function of its input, rather than deterministically, offers two practical advantages. The randomness makes the response tolerant of input noise, and it supports sampling-based inference in which a network represents a distribution through the rate of its spike trains \cite{cassidy2013}. A probabilistic neuron also fits the behaviour of emerging analogue synapses, whose conductance updates are inherently stochastic, so a digital neuron with a controlled random source provides a convenient interface to such devices.

Generating that randomness compactly is the central design problem. Stochastic computing represents a number as the probability that a bit in a stream is one, which lets multiplication and addition be performed by simple logic gates at the cost of longer evaluation times \cite{alaghi2013}. The pseudorandom source that drives a stochastic stream is most often a linear-feedback shift register (LFSR), because it produces a long, reproducible sequence with very little area and at the full clock rate \cite{lee2024design}. The same shift-register primitive underlies hardware probabilistic bits and other stochastic electronics that aim to compute with controlled noise \cite{camsari2017,thakur2016}. A neuron built from an LFSR, a small activation table, and an integrator therefore packages several of these established ideas into a single reusable block.

This paper is a detailed study of one such block. The neuron described here is also one of four blocks in an integrated neuromorphic suite that shares a common serial interface, reported separately \cite{suite2026}. The present paper goes deeper on the neuron than the suite paper does. It adds a stochastic characterisation of the random source and the comparator, a rate-coding and refractory study, and the single-block implementation figures, none of which the suite paper presents at this level of detail. The two papers are complementary: the suite paper describes the integrated four-block system and the shared interface, while this paper treats the neuron as a self-contained block and characterises its behaviour.

The contributions of this paper are the following. First, an open, configurable LFSR-stochastic LIF neuron on the SkyWater 130\,nm node, with a programmable feedback polynomial, an eight-entry activation table, a programmable threshold, a programmable decay, and a programmable refractory period. Second, a stochastic characterisation from a model checked bit-exact against the register-transfer description, covering the shift-register period as a function of the polynomial, the uniformity of the comparison value, the autocorrelation of the random source and of the comparator output including a serial-correlation property and its mitigation, and the per-entry firing probability. Third, a rate-coding and refractory characterisation that relates the output spike rate to the input weight, the threshold, and the refractory period. Fourth, an open SkyWater 130\,nm implementation with a routed layout and a directed verification suite. The techniques used here are established. Among these, the run-time configurable feedback polynomial, which most digital stochastic sources fix in hardware, and the serial-correlation characterisation of the overlapping-byte comparator are the most distinctive; the remainder is a careful open implementation of established parts. The contribution is an openly released, characterised implementation, not a new neuron model and not measured silicon.

The remainder of the paper is organised as follows. Section~\ref{sec:bg} reviews related silicon neurons and stochastic random sources. Section~\ref{sec:arch} describes the neuron architecture with its governing equations. Section~\ref{sec:stoch} reports the stochastic characterisation. Section~\ref{sec:rate} reports the rate-coding and refractory behaviour. Section~\ref{sec:impl} reports the implementation and verification. Section~\ref{sec:disc} discusses trade-offs and limitations, and Section~\ref{sec:concl} concludes.

\section{Background and related work}\label{sec:bg}
\textbf{Silicon neurons.} Hardware neurons span a wide range of implementation styles, from analogue circuits that emulate membrane dynamics directly to fully digital datapaths \cite{indiveri2011}. The leaky integrate-and-fire (LIF) model is the most common choice in spiking hardware because it captures temporal integration with a single state variable and a threshold comparison \cite{venkateswara2024effici}. Recent work has examined digital LIF datapaths optimised for leakage and event-driven efficiency \cite{krausse2024explor} and the dynamical behaviour of compact neuron circuits \cite{orima2024bifurc}. Large neurosynaptic cores use digital, pseudorandom-driven neurons with a refractory period to realise probabilistic firing at scale \cite{cassidy2013,merolla2014}, and the neuron in this work follows the same model: a deterministic integrator gated by a pseudorandom comparator.

\textbf{Stochastic computing and LFSR random sources.} Stochastic computing performs arithmetic on bit streams whose value is the probability of a one, which reduces complex operations to small logic at the cost of evaluation time \cite{alaghi2013}. The quality and cost of the pseudorandom source dominate the accuracy and area of such designs, and recent work has studied LFSR-based generation, its hardware quantisation, and its energy efficiency \cite{lee2024design,lee2025hwquant,saikia2024adapti,akter2025highen}. Random-number generation for stochastic and security applications continues to be an active topic \cite{ichikawa2025random}. Stochastic activation in a neuron, in which firing is a probabilistic function of the input, reduces sensitivity to input noise and supports sampling-based computation \cite{wang2014a2}. The present neuron uses an LFSR and a comparator against a programmable table to realise this activation.

\textbf{Probabilistic bits and stochastic electronics.} A related line of work computes with controlled randomness using probabilistic bits, in which a device fluctuates between two states with a tunable probability \cite{camsari2017}, and with other stochastic-electronic primitives intended for low-power inference \cite{thakur2016}. These approaches share the goal of replacing exact arithmetic with controlled noise, which is the same trade made by an LFSR-driven stochastic neuron in the digital domain.

\textbf{Plasticity context.} Spike-timing-dependent plasticity adjusts a synaptic weight from the relative timing of pre- and post-synaptic spikes and is a widely used unsupervised rule in spiking systems \cite{bipoo1998,mostafa2014a}. The neuron reported here does not implement learning; it produces the spike train that a downstream plasticity rule would consume, and learning is provided by a separate block in the companion suite \cite{suite2026}.

\textbf{Open physical design and Tiny Tapeout on SkyWater 130\,nm.} The SkyWater 130\,nm process and its open process design kit, together with the OpenLane and OpenROAD flows, have made standard-cell tape-out accessible for academic blocks \cite{shalan2020buildi,herman2023design}. Open neuromorphic hardware has appeared on this and similar flows, including open spiking accelerators on the SkyWater node \cite{modaresi2023}, open digital neuromorphic processors with on-chip learning \cite{frenkel2019,frenkel2022}, and open field-programmable emulators \cite{neurocorex2025}, and recent stochastic-computing neurons use the same ingredients of an LFSR and a comparator \cite{sclif2026}. The block was submitted through the Tiny Tapeout shared-silicon programme, which provides a standardised tile and pin harness. The present work is implemented on the same node and flow, which fixes a stable, well-documented target.

\section{Neuron architecture}\label{sec:arch}
\begin{figure}[tbp]
\centering
\includegraphics[width=0.92\textwidth]{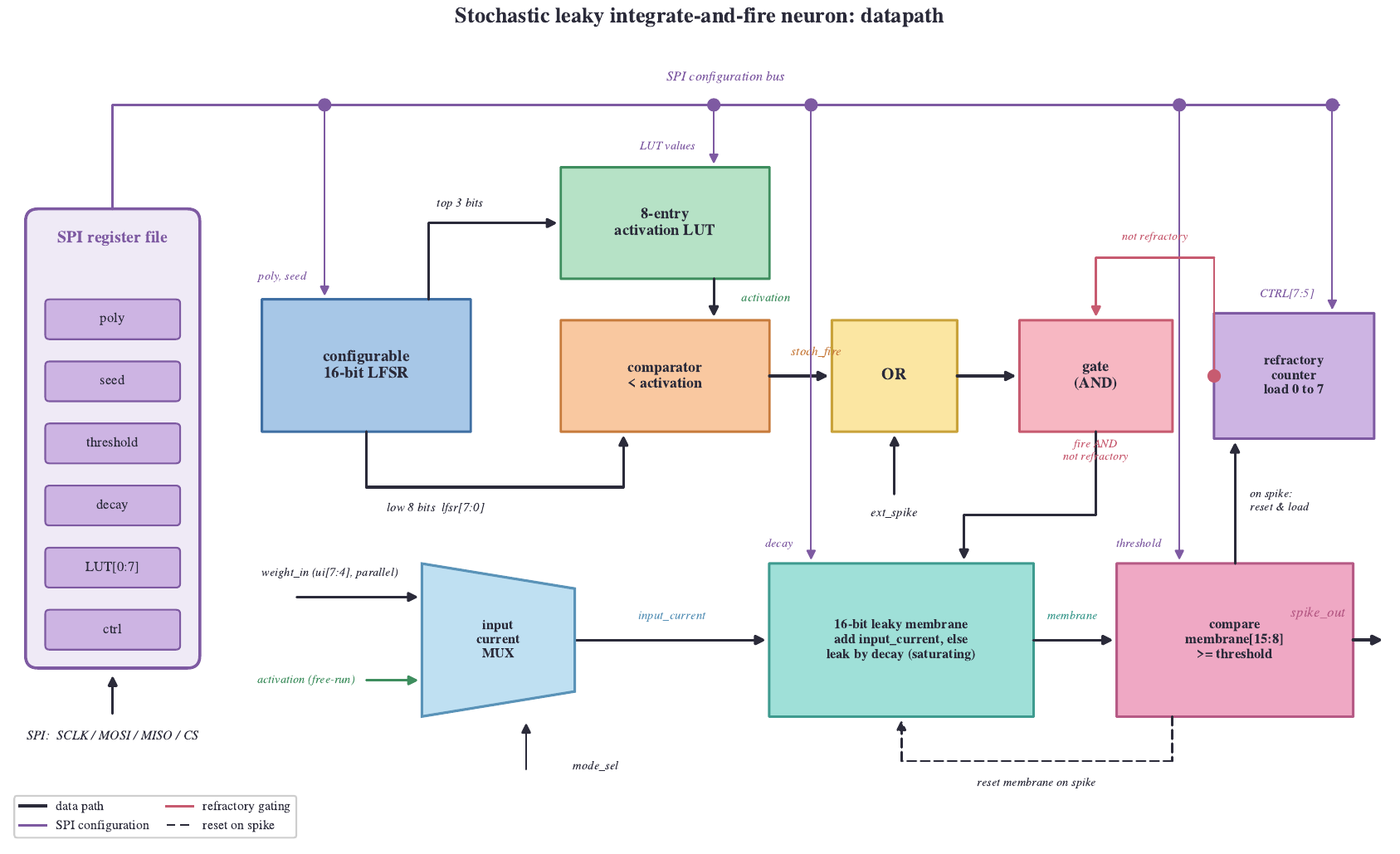}
\caption{Datapath of the stochastic leaky integrate-and-fire neuron. The configurable linear-feedback shift register supplies pseudorandom bits each cycle. The top three bits select one of eight activation-table entries, and the low eight bits form the comparison value; a stochastic event is asserted when the comparison value is less than the selected activation. The leaky integrate-and-fire core adds the input current on a stochastic event or an external spike, saturating at full scale, and otherwise leaks by the programmable decay. A spike is emitted when the upper byte of the membrane reaches the threshold, after which the membrane resets and the refractory counter loads. All parameters are written through the serial register file.}
\label{fig:arch}
\end{figure}

\begin{figure}[tbp]
\centering
\includegraphics[width=0.92\textwidth]{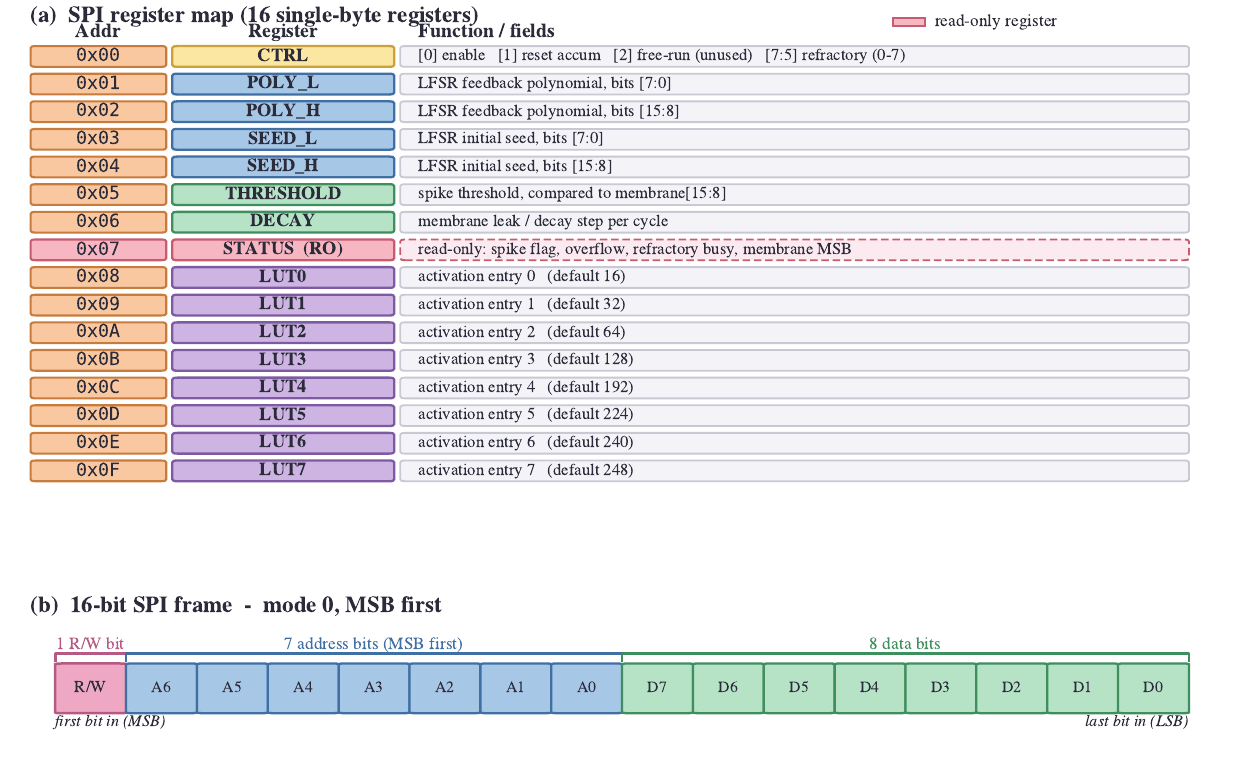}
\caption{Serial register interface. (a) The sixteen single-byte registers that configure the neuron, grouped by function: control and refractory, feedback polynomial, seed, threshold, decay, status, and the eight activation-table entries. (b) The sixteen-bit serial frame, with one read-or-write bit, seven address bits sent most significant first, and eight data bits.}
\label{fig:regmap}
\end{figure}

The neuron has three parts: a configurable LFSR that supplies pseudorandom bits, an activation table that converts those bits into a per-cycle firing probability, and a leaky integrate-and-fire core that converts probabilistic events into a spike train. Figure~\ref{fig:arch} shows the datapath, and figure~\ref{fig:lfsr} shows the shift register in detail. All parameters are set through the sixteen-register serial interface summarised in figure~\ref{fig:regmap} and Table~\ref{tab:regs}.

\subsection{Configurable linear-feedback shift register}
The random source is a 16-bit Fibonacci LFSR. Let $s_t$ denote the 16-bit state at cycle $t$ and let $p$ denote the 16-bit feedback polynomial held in a register. The feedback bit is the parity of the bitwise conjunction of the state and the polynomial, that is the exclusive-or reduction of the selected taps,
\begin{equation}
f_t \;=\; \bigoplus_{i=0}^{15} \bigl( s_t[i] \wedge p[i] \bigr),
\label{eq:feedback}
\end{equation}
where $\bigoplus$ denotes reduction by exclusive-or and $\wedge$ denotes bitwise conjunction. The next state is the current state shifted left by one position with the feedback bit inserted in the least significant bit,
\begin{equation}
s_{t+1} \;=\; \bigl( (s_t \ll 1) \;\vee\; f_t \bigr) \bmod 2^{16},
\label{eq:shift}
\end{equation}
where $\ll$ denotes a left shift and $\vee$ denotes bitwise disjunction. The all-zero state is absorbing for any linear feedback, so the logic detects a zero state and reseeds it to the value one, which keeps the register running. Both the polynomial $p$ and the initial seed are written through registers, so the period and the sequence are configurable at run time.

\subsection{Activation table and stochastic event}
The activation table has eight entries, each an eight-bit value. The top three bits of the state, $s_t[15{:}13]$, index the table and select the activation $a_t$ for the current cycle. The low eight bits of the state, $s_t[7{:}0]$, form an eight-bit comparison value $c_t$. A stochastic event is asserted when the comparison value is strictly less than the selected activation,
\begin{equation}
\mathrm{stoch\_fire}_t \;=\; \bigl[\, c_t < a_t \,\bigr].
\label{eq:stochfire}
\end{equation}
Because the comparison value is uniform over the eight-bit range across the full period, as shown in Section~\ref{sec:stoch}, the probability of a stochastic event for a selected entry is the entry value divided by 256,
\begin{equation}
\Pr[\mathrm{stoch\_fire} \mid a] \;=\; \frac{a}{256}.
\label{eq:prob}
\end{equation}
The eight-entry table therefore acts as a programmable, piecewise activation function mapped onto the firing probability.

\subsection{Leaky integrate-and-fire core}
The membrane is a 16-bit unsigned accumulator $v_t$. On a stochastic event or an external input spike, the input current $I$ is added to the membrane, saturating at the 16-bit full scale; otherwise the membrane decays by a programmable amount $d$, floored at zero,
\begin{equation}
v_{t+1} \;=\;
\begin{cases}
\min\bigl(v_t + I,\; 2^{16}-1\bigr), & \text{event or input spike},\\[2pt]
\max\bigl(v_t - d,\; 0\bigr), & \text{otherwise}.
\end{cases}
\label{eq:membrane}
\end{equation}
A spike is emitted when the upper byte of the membrane reaches the threshold $\theta$, that is when $v_t[15{:}8] \geq \theta$. On a spike the membrane is reset and a refractory counter is loaded with the programmed value $r$, which lies in the range zero to seven. While the refractory counter is non-zero it decrements each cycle and suppresses both integration and firing, so the membrane is held until the counter expires.

The block operates in two modes selected by an input pin (\code{ui\_in[1]}). In free-run mode the input current equals the selected activation, $I = a_t$, so the same table value sets both the firing probability and the integration step. In host mode the input current is taken instead from a four-bit weight presented on parallel input pins (\code{ui\_in[7:4]}); the serial interface still configures the polynomial, table, threshold, decay, and refractory period and can enable the block, but the per-cycle weight is a parallel input rather than a register. In both modes the integrator, threshold, decay, and refractory logic are identical.

The neuron exposes its internal state for observation and bring-up. The single-bit spike output is brought out directly, and the most significant bit of the shift register is brought out as a low-cost randomness monitor, which is the signal sampled for the autocorrelation in Section~\ref{sec:stoch}. The upper four bits of the membrane are brought out as a slow, analogue-like bus that a bench instrument or an on-chip converter can read as a state-of-health signal, and an accumulator-overflow flag and a latched-spike flag are exposed alongside it. An external spike input on a parallel pin can drive the integrator directly, in addition to the stochastic path, and a read-only register returns the live refractory-counter value. These observability and interfacing points make the block straightforward to bring up on returned silicon and let it act as a source of events for a downstream synapse or router.

\begin{table}[tbp]
\centering
\caption{Serial register map of the stochastic neuron. All sixteen registers are single bytes. Access is read-write except where marked read-only (RO). The polynomial and seed are split across a low and a high byte, and the eight activation-table entries occupy addresses \code{0x08} to \code{0x0F}.}
\label{tab:regs}
\small
\begin{tabular}{@{}p{1.4cm}p{2.4cm}p{1.4cm}p{7.6cm}@{}}
\toprule
\rowcolor{headpastel}
\textbf{Address} & \textbf{Name} & \textbf{Access} & \textbf{Description}\\
\midrule
\code{0x00} & CTRL & R/W & [0] enable, [1] accumulator reset, [2] free-run, [7:5] refractory period (0 to 7)\\
\rowcolor{rowpastel}
\code{0x01} & POLY\_L & R/W & LFSR feedback polynomial, bits [7:0]\\
\code{0x02} & POLY\_H & R/W & LFSR feedback polynomial, bits [15:8]\\
\rowcolor{rowpastel}
\code{0x03} & SEED\_L & R/W & LFSR initial seed, bits [7:0]\\
\code{0x04} & SEED\_H & R/W & LFSR initial seed, bits [15:8]\\
\rowcolor{rowpastel}
\code{0x05} & THRESHOLD & R/W & Spike threshold, compared with the membrane upper byte [15:8]\\
\code{0x06} & DECAY & R/W & Membrane leak step per cycle\\
\rowcolor{rowpastel}
\code{0x07} & STATUS & RO & Spike flag, accumulator overflow, latched spike, membrane upper bit, refractory busy\\
\code{0x08} & LUT0 & R/W & Activation entry 0\\
\rowcolor{rowpastel}
\code{0x09} & LUT1 & R/W & Activation entry 1\\
\code{0x0A} & LUT2 & R/W & Activation entry 2\\
\rowcolor{rowpastel}
\code{0x0B} & LUT3 & R/W & Activation entry 3\\
\code{0x0C} & LUT4 & R/W & Activation entry 4\\
\rowcolor{rowpastel}
\code{0x0D} & LUT5 & R/W & Activation entry 5\\
\code{0x0E} & LUT6 & R/W & Activation entry 6\\
\rowcolor{rowpastel}
\code{0x0F} & LUT7 & R/W & Activation entry 7\\
\bottomrule
\end{tabular}
\end{table}

\section{Stochastic characterisation}\label{sec:stoch}
This section characterises the random source and the comparator from a model checked bit-exact against the register-transfer code and the gate-level simulation. The model reproduces the LFSR state sequence, the comparison value, and the stochastic-event signal cycle by cycle, which makes the statistics below properties of the implemented logic rather than of an idealised generator.

\subsection{Period versus polynomial}
\begin{figure}[tbp]
\centering
\includegraphics[width=0.92\textwidth]{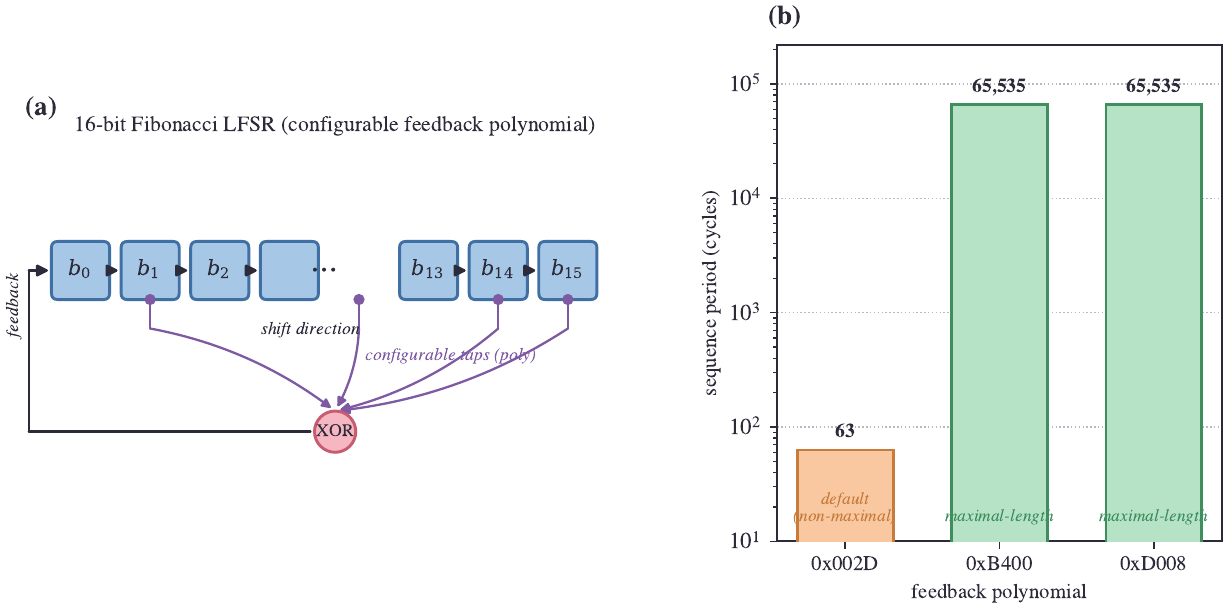}
\caption{Configurable linear-feedback shift register. (a) Structure of the 16-bit Fibonacci register: the feedback bit is the exclusive-or reduction of the bitwise conjunction of the state and the programmable polynomial, and the next state is the state shifted left with the feedback inserted in the least significant bit. (b) Measured period as a function of the feedback polynomial. The shipped default \code{0x002D} produces a short cycle of 63 states, while maximal-length polynomials such as \code{0xB400} and \code{0xD008} produce the full period of 65535 states.}
\label{fig:lfsr}
\end{figure}

The period of the shift register depends on the feedback polynomial. The polynomial that ships as the default, \code{0x002D}, is not a maximal-length polynomial; it yields a short cycle of 63 states, and the reset seed lies on a 10-state transient that feeds into that cycle. Maximal-length polynomials, such as \code{0xB400} and \code{0xD008}, yield the full period of $2^{16}-1 = 65535$ states. Figure~\ref{fig:lfsr}(b) shows the measured period for these polynomials. The polynomial is programmable, so for full-period operation a maximal-length value should be programmed into the polynomial register before use; the short default is suitable only where a short, repeating sequence is acceptable.

\subsection{Uniformity of the comparison value}
\begin{figure}[tbp]
\centering
\includegraphics[width=0.92\textwidth]{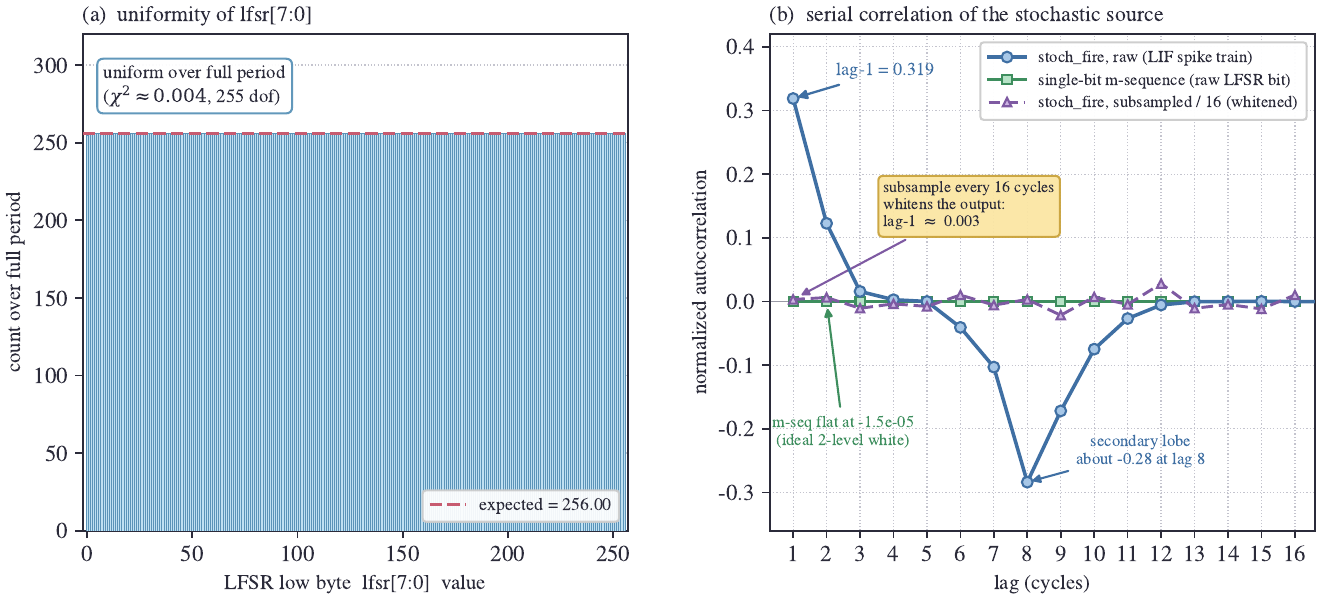}
\caption{Statistics of the random source over the full maximal-length period (polynomial 0xB400). (a) Histogram of the eight-bit comparison value; each value occurs the same number of times to the resolution of the count, so the comparison value is uniform. (b) Autocorrelation of the stochastic-event signal as a function of lag. The signal is serially correlated, about $0.32$ at lag one with a secondary negative lobe of about $-0.28$ near lag eight, because the compared byte shifts by one bit each cycle and consecutive bytes share seven of eight bits; subsampling the output every sixteen cycles suppresses the lobe and reduces the lag-one autocorrelation to near zero.}
\label{fig:random}
\end{figure}

Over the full maximal-length period the eight-bit comparison value $c_t$ is uniform. Each of the 256 possible byte values appears 256 times across the period, with the single exception that follows from excluding the all-zero state, which removes one occurrence. Figure~\ref{fig:random}(a) shows the histogram of the comparison value over the full period; the distribution is flat to the resolution of the count. This uniformity is the property that makes the firing probability in equation~\eqref{eq:prob} equal to the activation divided by 256.

\subsection{Autocorrelation and a serial-correlation property}
The single-bit maximal-length sequence produced by the shift register has the ideal two-level autocorrelation of a maximal-length sequence: after the usual mapping of the bit to plus or minus one, its unnormalised periodic autocorrelation is constant and equal to $-1$ for all non-zero lags, which once normalised by the period $N = 65535$ becomes $-1/N \approx -1.5 \times 10^{-5}$, indistinguishable from zero on the normalised scale of Figure~\ref{fig:random}(b). The comparator output, however, is not white at short lags. The comparison value $c_t$ is formed from eight consecutive bits of the register, and because the register shifts by one bit each cycle, the byte at cycle $t+1$ shares seven of its eight bits with the byte at cycle $t$. Consecutive comparison values are therefore strongly overlapping, and the stochastic-event signal inherits that overlap as serial correlation. The measured autocorrelation of the stochastic-event signal is about $0.32$ at lag one and falls through the first few lags, but it does not remain near zero: a secondary negative lobe of about $-0.28$ appears near lag eight, which reflects the eight-bit width of the comparison value together with the one-bit-per-cycle shift, before the autocorrelation decays again at longer lags.

This kind of correlation between overlapping linear-feedback shift register outputs is recognised in stochastic computing, where the seeding, scrambling, and feedback polynomial all affect stream quality \cite{anderson2016lfsr}; it is reported here as a concrete property of this comparator rather than as a new effect, and it is the kind of property that a system-level model treating the register as an ideal source would not expose. A simple mitigation removes it: subsampling the stochastic-event signal every sixteen cycles, so that the bytes used no longer overlap and the negative lobe at lag eight is cleared, reduces the lag-one autocorrelation to about $0.003$, which is near whiteness. A stride of eight is not sufficient, because the subsampled lag-one autocorrelation then equals the raw autocorrelation at lag eight, about $-0.28$; a stride of sixteen is the smallest power-of-two stride that clears both the positive short-lag correlation and the negative lobe. Figure~\ref{fig:random}(b) shows the raw, the subsampled, and the ideal single-bit autocorrelation. The mitigation costs throughput, because only one cycle in sixteen is used, which at the 50\,MHz clock corresponds to a maximum near-independent sample rate of about 3.125 million samples per second; it is appropriate where the downstream computation requires near-independent samples.

\subsection{Per-entry firing probability}
\begin{figure}[tbp]
\centering
\includegraphics[width=0.78\textwidth]{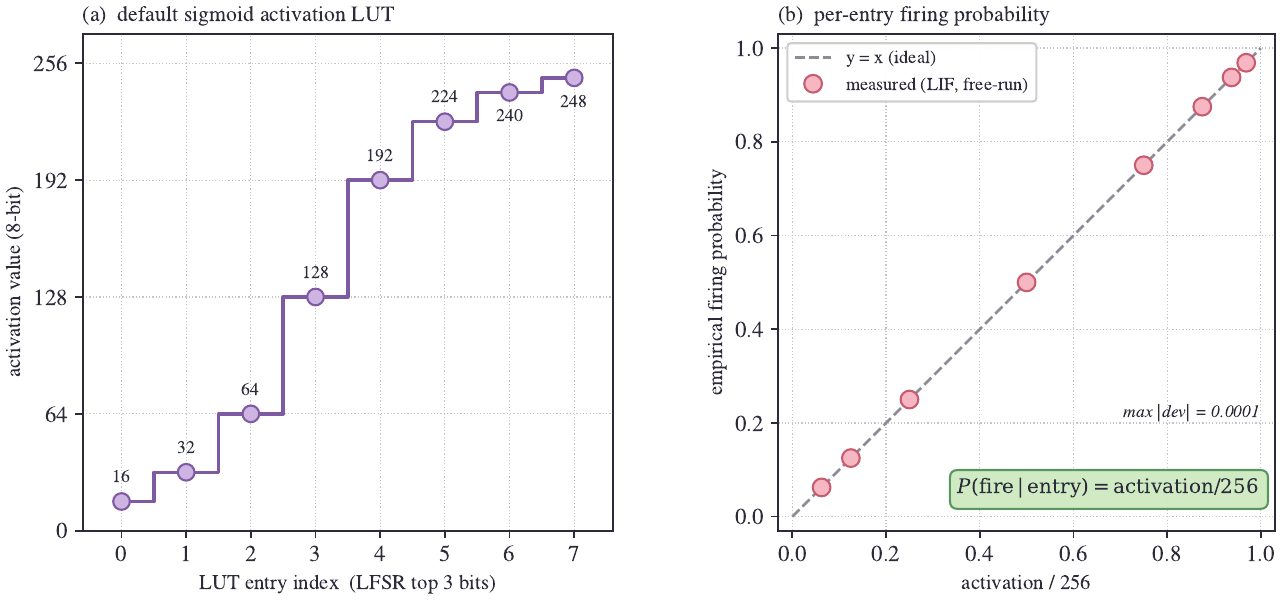}
\caption{Activation table and firing probability. (a) The default activation table: the eight programmed entries, addressed by the top three shift-register bits, rise in eight-bit steps and approximate a sigmoid. (b) Measured per-entry firing probability against the activation value. The points are the fraction of cycles on which a stochastic event is asserted, counted over the full maximal-length period; the line is the relation $a/256$ from equation~\eqref{eq:prob}, so the activation table sets a Bernoulli firing probability with eight-bit resolution. The sweep uses the maximal-length polynomial 0xB400 over the full period with the default activation table.}
\label{fig:activation}
\end{figure}

The firing probability of each table entry follows equation~\eqref{eq:prob}. Sweeping the activation value over the eight-entry table and counting stochastic events over the full period gives a measured firing probability equal to the activation divided by 256 at each entry, to the resolution of the count. Figure~\ref{fig:activation} shows the measured firing probability against the activation value, together with the line $a/256$. The agreement confirms that the activation table sets a Bernoulli firing probability with eight-bit resolution, and that the model used for the statistics in this section matches the register-transfer and gate-level behaviour. All of the statistics in this section are reported for a maximal-length polynomial; with the shipped default polynomial the sequence repeats every 63 states, none of these properties hold, and a maximal-length value must be programmed for full-period operation.

\section{Rate coding and dynamics}\label{sec:rate}
The neuron encodes its input as an output spike rate. Each stochastic event or input spike adds the input current to the membrane, the membrane crosses the threshold after a number of events that depends on the current and the threshold, and the refractory period limits how often a spike can follow another. The output rate is therefore a monotonic function of the input weight and of the threshold, with a hard cap set by the refractory period.

\subsection{Control of the output rate}
\begin{figure}[tbp]
\centering
\includegraphics[width=0.92\textwidth]{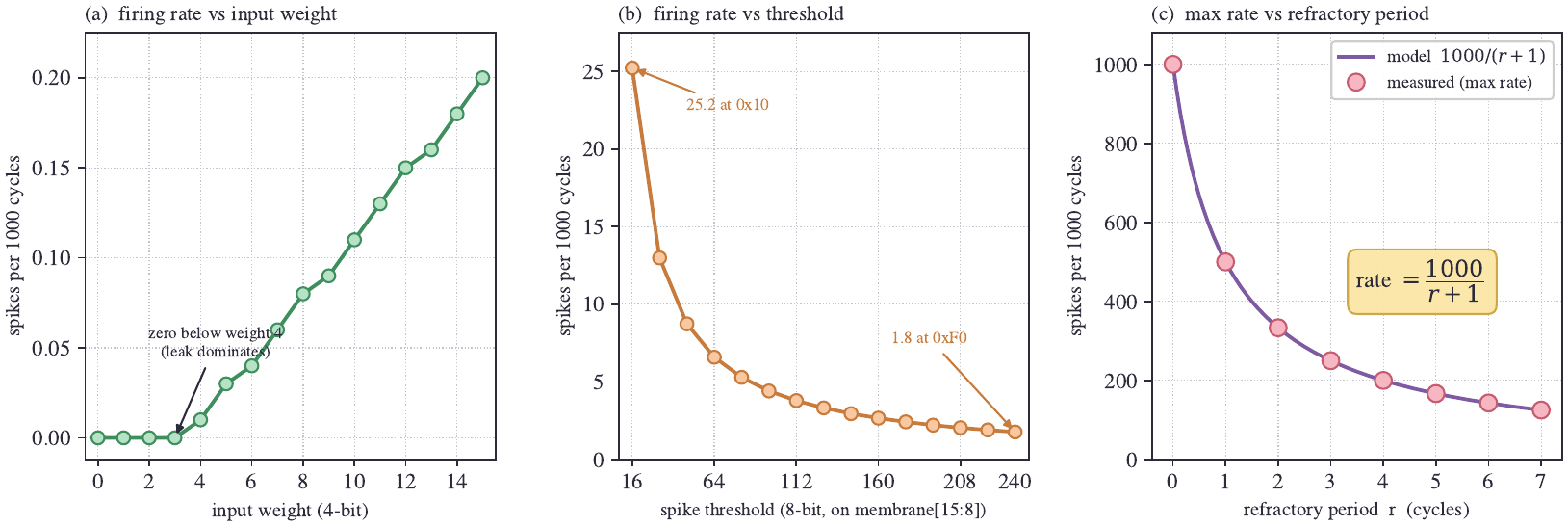}
\caption{Rate-coding and refractory behaviour. The output spike rate increases monotonically with the input weight and decreases monotonically with the threshold. The refractory period caps the maximum output rate; the measured maximum-rate points lie on the law $1/(r+1)$ spikes per cycle, where $r$ is the programmed refractory value in the range zero to seven. The weight sweep uses host mode with threshold 0x80 and decay 4, the threshold sweep uses the free-run stochastic drive with decay 4, and the refractory sweep applies an external spike every cycle with the threshold at its minimum so that the rate is refractory-limited; all sweeps use the maximal-length polynomial 0xB400.}
\label{fig:tuning}
\end{figure}

The output spike rate increases monotonically with the input weight, because a larger current advances the membrane toward the threshold in fewer events. The output rate decreases monotonically with the threshold, because a higher threshold requires more accumulation before the upper byte of the membrane reaches it. Figure~\ref{fig:tuning} shows the output rate as a function of the input weight and as a function of the threshold; both sweeps are monotonic over the configured range. These two controls, the weight and the threshold, give a host two independent ways to set the operating point of the neuron.

\subsection{Refractory cap and membrane behaviour}
The refractory period sets an upper bound on the output rate. After a spike, the refractory counter loads with the programmed value $r$ and suppresses both integration and firing while it is non-zero, so the soonest a further spike can occur is $r+1$ cycles after the previous spike. The maximum output rate is therefore $1/(r+1)$ spikes per cycle. The measured maximum-rate points lie on this law across the range $r = 0$ to $r = 7$, as shown in figure~\ref{fig:tuning}. Between spikes the membrane ramps upward as events accumulate, crosses the threshold, resets to zero, and is then held by the refractory counter until the counter expires, after which integration resumes. When no events arrive the membrane leaks toward zero by the programmable decay, which sets the time over which past input is forgotten.

\section{Implementation and verification}\label{sec:impl}
The neuron was taken through an open standard-cell flow using OpenLane and OpenROAD on the SkyWater 130\,nm process. It is mapped to a single Tiny Tapeout 1x1 tile of the target shuttle, about 161\,$\mu$m by 112\,$\mu$m, and is constrained to a 50\,MHz system clock. Table~\ref{tab:ppa} summarises the implementation figures from synthesis and place-and-route.

\begin{table}[tbp]
\centering
\caption{Implementation summary for the stochastic neuron on SkyWater 130\,nm. All figures are from the open synthesis and place-and-route flow at a 50\,MHz constraint. The power figure is a pre-silicon estimate after clock-tree synthesis at a default switching activity and is not a measurement.}
\label{tab:ppa}
\small
\begin{tabular}{@{}p{6.8cm}p{6.2cm}@{}}
\toprule
\rowcolor{headpastel}
\textbf{Metric} & \textbf{Value}\\
\midrule
Process & SkyWater 130\,nm open process design kit\\
\rowcolor{rowpastel}
Tile & Single Tiny Tapeout 1x1, about 161\,$\mu$m by 112\,$\mu$m\\
Clock constraint & 50\,MHz\\
\rowcolor{rowpastel}
Standard cells & 917\\
Flip-flops & 208 (179 dfrtp and 29 dfstp)\\
\rowcolor{rowpastel}
Post-synthesis standard-cell area & about 10{,}593 square micrometres\\
Placed utilisation & 70 per cent\\
\rowcolor{rowpastel}
Worst setup slack & about $+14.3$\,ns at 50\,MHz\\
Worst hold slack & about $+0.35$\,ns at 50\,MHz\\
\rowcolor{rowpastel}
Estimated power & about 701\,$\mu$W post-clock-tree-synthesis, default activity (pre-silicon estimate)\\
Verification & eighteen directed cocotb tests at register-transfer and gate level, all passing\\
\bottomrule
\end{tabular}
\end{table}

\subsection{Physical implementation}
\begin{figure}[tbp]
\centering
\includegraphics[width=0.62\textwidth]{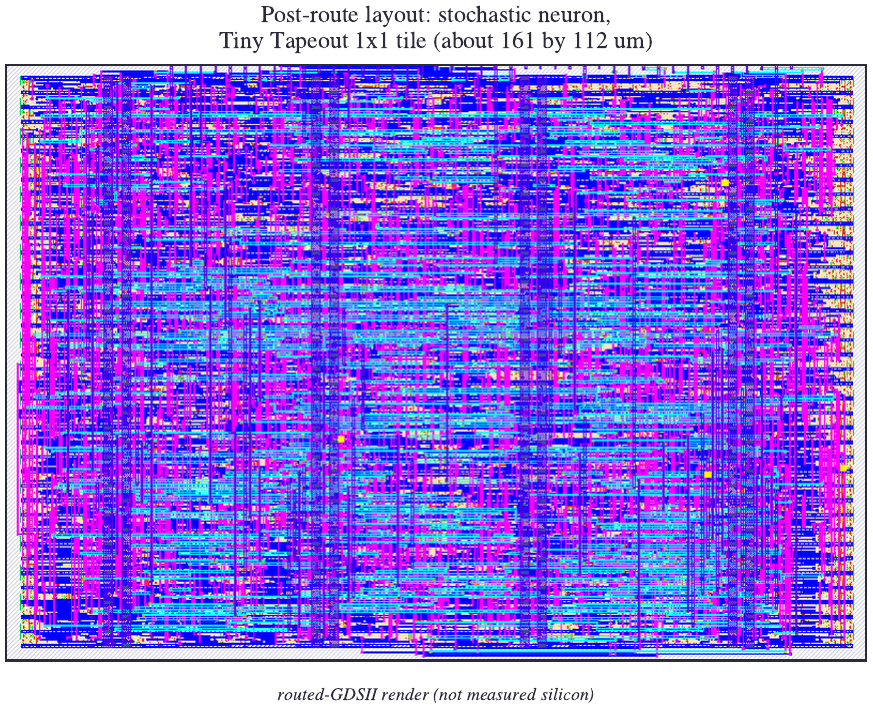}
\caption{Routed layout of the stochastic neuron on a single Tiny Tapeout 1x1 tile on the SkyWater 130\,nm process. The image is a render from the open place-and-route flow and is not measured silicon.}
\label{fig:layout}
\end{figure}

The mapped design uses 917 standard cells, of which 208 are flip-flops (179 of type dfrtp and 29 of type dfstp). The post-synthesis standard-cell area is about 10{,}593 square micrometres, and the design places at 70 per cent utilisation on the single tile. Static timing analysis at the 50\,MHz constraint reports a worst setup slack of about $+14.3$\,ns and a worst hold slack of about $+0.35$\,ns, so the design meets timing with positive margin on both checks. The estimated power after clock-tree synthesis is about 701\,$\mu$W at a default switching-activity assumption; this is a pre-silicon estimate from the flow, not a measurement. The largest combinational cell type is the multiplexer, which reflects the sixteen-register file and the read multiplexing of the eight-entry activation table. Figure~\ref{fig:layout} shows the routed layout of the tile; the image is a render from the implementation flow and is not measured silicon.

\subsection{Verification}
The block is verified by eighteen directed cocotb test cases exercised at register-transfer and gate level, all passing. The tests cover the serial register interface, the configurable polynomial and seed, the activation table and the stochastic-event comparison, the integrate-and-fire dynamics including saturation and decay, the threshold comparison on the upper byte, and the refractory counter across its range. The same model used for the statistics in Section~\ref{sec:stoch} is checked bit-exact against these simulations, which links the reported statistics to the verified logic.

\subsection{Qualitative comparison}
Table~\ref{tab:compare} places the neuron alongside a large digital neurosynaptic core and a recent stochastic-computing neuron. The comparison is qualitative and lists the technology node, the neuron and datapath style, the randomness source, the on-chip programmability, whether the design is open-source, and whether the figures are from measured silicon or are pre-silicon. The present work is marked pre-silicon throughout, because no fabricated device is reported.

\begin{table}[tbp]
\centering
\caption{Qualitative comparison with a digital neurosynaptic core and a stochastic-computing neuron. The comparison is on design attributes, not measured performance. The present work is pre-silicon; the entries for the other designs are taken from their published descriptions.}
\label{tab:compare}
\footnotesize
\begin{tabular}{@{}p{3.0cm}p{2.4cm}p{2.5cm}p{2.0cm}p{2.7cm}@{}}
\toprule
\rowcolor{headpastel}
\textbf{Attribute} & \textbf{TrueNorth digital neuron \cite{cassidy2013,merolla2014}} & \textbf{LFSR stochastic neuron \cite{sclif2026}} & \textbf{Recent LFSR design \cite{lee2025hwquant}} & \textbf{This work}\\
\midrule
Node & 28\,nm CMOS & open flow & technology study & SkyWater 130\,nm\\
\rowcolor{rowpastel}
Neuron / datapath & digital LIF, pseudorandom & stochastic LIF & stochastic-computing datapath & stochastic LIF, saturating integrator\\
Randomness source & pseudorandom generator & LFSR & LFSR & configurable-polynomial LFSR\\
\rowcolor{rowpastel}
On-chip programmability & neuron parameters & threshold and table & quantisation settings & polynomial, table, threshold, decay, refractory\\
Open-source & no & yes & not stated & yes\\
\rowcolor{rowpastel}
Silicon & measured & pre-silicon & study & pre-silicon\\
\bottomrule
\end{tabular}
\end{table}

\section{Discussion}\label{sec:disc}
\textbf{Design trade-offs.} Several choices in the neuron trade resolution or quality for area and timing. The threshold comparison uses only the upper byte of the 16-bit membrane, which narrows the comparison path and eases timing, at the cost of reducing the effective threshold resolution to 256 levels. The activation table has eight entries, which keeps the table small and the index path short, at the cost of a coarse, piecewise activation function; finer activation would require a larger table or interpolation. The most interesting trade is in the random source. Reading eight consecutive register bits as the comparison value is the lowest-area way to obtain a byte each cycle, but it makes consecutive comparison values overlap and produces the serial correlation reported in Section~\ref{sec:stoch}. The subsampling mitigation restores near-whiteness at the cost of throughput, so the choice between the raw output and the subsampled output depends on whether the downstream computation needs near-independent samples or can tolerate short-lag correlation. A final caveat concerns the default polynomial, which is not maximal-length; a maximal-length value should be programmed where a full-period sequence is required.

\textbf{Limitations.} The results in this paper are from simulation and the open implementation flow. The reported area, timing, and power are pre-silicon, and the power figure in particular depends on a default switching-activity assumption and must be confirmed by measurement on the returned chip. The quality of the random source, including the serial-correlation property and the effect of subsampling, is characterised from the bit-exact model and should likewise be confirmed on silicon. The block is a single neuron, not an array, so throughput and area for a population of neurons are outside the scope of this study.

\textbf{Relation to the suite paper.} This paper is the single-block study of the neuron. The companion paper describes the integrated four-block suite, which places the neuron alongside a ring-oscillator sensor, a spike-timing-dependent plasticity controller, and a crossbar controller behind one shared serial interface \cite{suite2026}. The suite paper covers the shared interface and the system-level composition; the present paper covers the neuron in the detail that a single-block study allows, including the stochastic characterisation and the rate-coding study that the suite paper does not present.

\textbf{Path to silicon.} The block has been submitted through the Tiny Tapeout shared-silicon programme. On return, the priorities are to measure the power under realistic switching activity, to confirm the timing margins reported by static analysis, and to verify the randomness quality of the comparator output and the effect of the subsampling mitigation. These measurements would convert the pre-silicon figures in this paper into measured results.

\FloatBarrier
\section{Conclusion}\label{sec:concl}
This paper presented a compact, configurable stochastic leaky integrate-and-fire neuron on the SkyWater 130\,nm process, implemented in standard cells and released openly. The neuron combines a configurable-polynomial linear-feedback shift register, an eight-entry programmable activation table, and a saturating leaky integrator with a programmable threshold and refractory period, all set through a sixteen-register serial interface. A characterisation from a model checked bit-exact against the register-transfer code established the period as a function of the polynomial, the uniformity of the comparison value, the per-entry firing probability of activation divided by 256, and a serial-correlation property of the comparator output with a subsampling mitigation. Rate-coding sweeps showed monotonic control of the output rate by the input weight and the threshold, with the refractory period capping the rate at one spike per refractory-plus-one cycles. The block occupies about 10,600 square micrometres of standard cells, places at 70 per cent utilisation on a single Tiny Tapeout tile, meets the 50\,MHz constraint with positive setup and hold margin, and passes eighteen directed cocotb tests. All results are from simulation and the open implementation flow, and no fabricated silicon is reported. The work is an openly released companion to a four-block neuromorphic suite reported separately.

\section*{Data, code, and reproducibility}
The register-transfer source, the cocotb test benches, the implementation configuration, and the design documentation for the neuron are public at \url{https://github.com/santhoshs93/tt_um_santhosh_stoch_neuron}, pinned to commit 225ce6e. The repositories for the four-block suite are described in the companion paper \cite{suite2026}.

The register-transfer behaviour reproduces directly: with cocotb and Icarus Verilog installed, \code{make -B} in the \code{test} directory runs the eighteen directed tests of Section~\ref{sec:impl}, and the bit-exact reference model behind the Section~\ref{sec:stoch} statistics follows from the same public source.

Table~\ref{tab:ppa} and the gate-level test pass come from the open hardening flow rather than from stored files. Running the project's \code{gds} GitHub Actions workflow, or a local OpenLane run, on the pinned commit regenerates the gate-level netlist, the area and cell-count report, the timing report, and the post-clock-tree power estimate of Table~\ref{tab:ppa}; that netlist, copied into \code{test}, runs the cocotb suite at gate level with \code{make -B GATES=yes}. The flow is deterministic at fixed tool versions. The netlist, the OpenLane reports, and the routed layout for commit 225ce6e are archived in a tagged release (\code{v1.0-paper}), with an archival identifier to be added on deposit.

\section*{Author contributions}
Poornima Kumaresan: Data curation, Formal analysis, Visualization, Writing (review and editing). Santhosh Sivasubramani: Conceptualization, Methodology, Software, Investigation, Validation, Writing (original draft), Supervision, Project administration, Funding acquisition.

\section*{Funding}
This work was supported by the INTRINSIC Lab, Centre for Sensors, Instrumentation and Cyber-Physical System Engineering (SeNSE), Indian Institute of Technology Delhi, and by the IM00002G\_RB\_SG IoE Fund Grant (NFSG), Indian Institute of Technology Delhi.

\section*{Conflict of interest}
The authors declare no conflict of interest.

\section*{Acknowledgements}
The authors acknowledge the SkyWater 130\,nm process design kit, the open standard-cell implementation flow, and the Tiny Tapeout shared-silicon shuttle.

\clearpage
{\footnotesize
\bibliographystyle{ncestyle}
\bibliography{references}
}

\end{document}